\documentclass[aip,amsmath,amssymb,reprint,]{revtex4-1}
\usepackage{graphicx}
\usepackage{dcolumn}
\usepackage{bm}
\usepackage[utf8]{inputenc}
\usepackage[T1]{fontenc}
\usepackage{mathptmx}
\usepackage{etoolbox}
\usepackage{multirow}
\usepackage{hyperref}
\usepackage{upgreek}
\usepackage{xcolor}

\makeatletter
\def\@email#1#2{%
 \endgroup
 \patchcmd{\titleblock@produce}
  {\frontmatter@RRAPformat}
  {\frontmatter@RRAPformat{\produce@RRAP{*#1\href{mailto:#2}{#2}}}\frontmatter@RRAPformat}
  {}{}
}%
\makeatother
\begin{document}

\preprint{AIP/123-QED}

\title[Do cities have a unique magnetic pulse?]{Do cities have a unique magnetic pulse?}

\author{V. Dumont}
\email{vincentdumont11@gmail.com}
\affiliation{Computational Research Division, Lawrence Berkeley National Laboratory, Berkeley CA 94720 USA}
\author{T. A. Bowen}
\affiliation{Department of Physics, University of California, Berkeley California 94720-7300 USA}
\affiliation{Space Sciences Laboratory, University of California, Berkeley California 94720-7300 USA}
\author{R. Roglans}
\affiliation{Department of Physics, University of California, Berkeley California 94720-7300 USA}
\affiliation{Space Sciences Laboratory, University of California, Berkeley California 94720-7300 USA}
\author{G. Dobler}
\affiliation{Biden School of Public Policy and Administration, University of Delaware, Newark DE 19716 USA}
\affiliation{Department of Physics and Astronomy, University of Delaware, Newark DE 19716 USA}
\affiliation{Data Science Institute, University of Delaware, Newark DE 19713 USA}
\affiliation{Center for Urban Science and Progress, New York University, Brooklyn NY 11201 USA}
\author{M. S. Sharma}
\affiliation{Center for Urban Science and Progress, New York University, Brooklyn NY 11201 USA}
\author{A. Karpf}
\affiliation{{Civil and Urban Engineering, Tandon School of Engineering, New York University, Brooklyn NY 11201 USA}}
\author{S. D. Bale}
\affiliation{Department of Physics, University of California, Berkeley California 94720-7300 USA}
\affiliation{Space Sciences Laboratory, University of California, Berkeley California 94720-7300 USA}
\author{A. Wickenbrock}
\affiliation{Institut f\"ur Physik, Universit\"at Mainz, Staudingerweg 7, 55128 Mainz, Germany}
\affiliation{Helmholtz Institut Mainz, Staudingerweg 18, 55128 Mainz, Germany}
\author{E. Zhivun}
\affiliation{Department of Physics, University of California, Berkeley California 94720-7300 USA}
\author{T. Kornack}
\affiliation{Twinleaf LLC, 300 Deer Creek Drive, Plainsboro, NJ 08536, USA}
\author{J. S. Wurtele}
\affiliation{Department of Physics, University of California, Berkeley California 94720-7300 USA}
\author{D. Budker}
\affiliation{Department of Physics, University of California, Berkeley California 94720-7300 USA}
\affiliation{Johannes Gutenberg-Universit\"{a}t Mainz, 55128 Mainz, Germany}
\affiliation{Helmholtz Institut Mainz, 55128 Mainz, Germany}

\date{\today}

\begin{abstract}
We present a comparative analysis of urban magnetic fields between two American cities: Berkeley (California) and Brooklyn Borough of New York City (New York). Our analysis uses data taken over a four-week period during which magnetic field data were continuously recorded using a fluxgate magnetometer of 70\,pT/$\sqrt{\mathrm{Hz}}$ sensitivity. We identified significant differences in the magnetic signatures. In particular, we noticed that Berkeley reaches a near-zero magnetic field activity at night whereas magnetic activity in Brooklyn continues during nighttime. We also present auxiliary measurements acquired using magnetoresistive vector magnetometers (VMR), with sensitivity of 300\,pT/$\sqrt{\mathrm{Hz}}$, and demonstrate how cross-correlation, and frequency-domain analysis, combined with data filtering can be used to extract urban magnetometry signals and study local anthropogenic activities. Finally, we discuss the potential of using magnetometer networks to characterize the global magnetic field of cities and give directions for future development.
\end{abstract}

\maketitle

\begin{figure*}[t]
\includegraphics[width=0.98\textwidth]{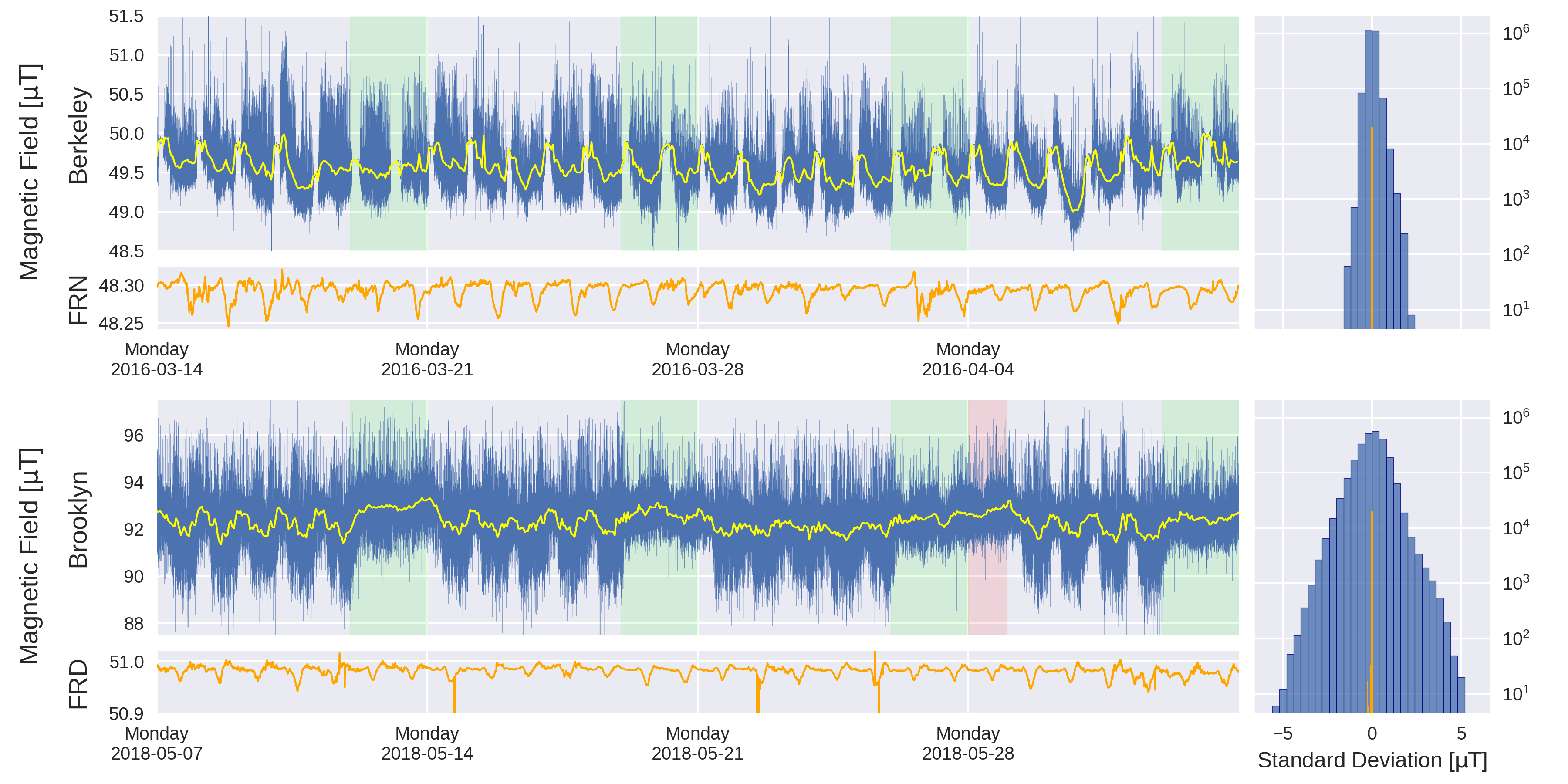}
\caption{Full 4-week time series of urban magnetic field data for Berkeley (top) and Brooklyn (bottom). Data downsampled to 1\,Hz are shown in blue while the data in yellow represent the downsampled hour-rate time series. The weekends are highlighted by the light green regions and holiday (Memorial Day) in light red. We note the differences in vertical scales between both cities; in particular, the excursions of magnetic field are significantly larger in Brooklyn. The geomagnetic field taken from the closest USGS station is shown in orange. The relative variation of the magnetic field around its mean value is shown on the right hand side with the distribution from the geomagnetic field represented in orange.}
\label{Fig:full_time_series}
\end{figure*}

\begin{figure*}[t]
\includegraphics[width=0.98\textwidth]{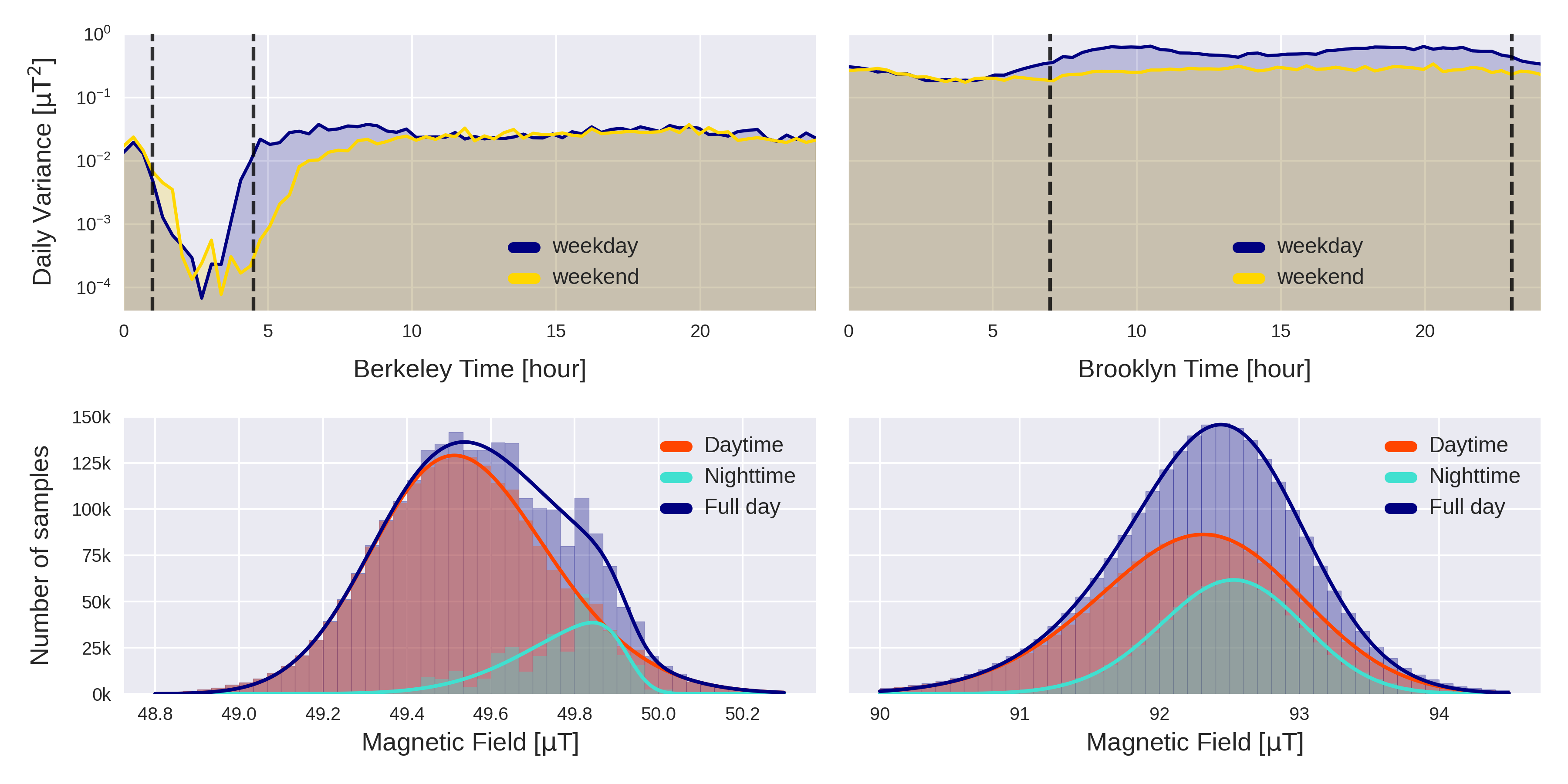}
\caption{Variance and distribution from Berkeley (left) and Brooklyn (right) data. The daily average variance was calculated over all the days for each consecutive 20-min time series. The vertical dashed lines on the top figures highlight the transition between day and night times. Nighttime has been set from 1 to 4:30\,AM for Berkeley and 11\,PM to 7\,AM for Brooklyn. The bottom plots show the daytime (red) and nighttime (green) distributions as well as for the full day (blue). A skewed Gaussian was fitted to both daytime and nighttime histograms independently (see Table\,\ref{table:best_fit} for best fit results). The blue skewed Gaussian profiles represent the sum of both daytime and nighttime profiles.}
\label{Fig:distcomp}
\end{figure*}

\begin{figure*}[t]
\includegraphics[width=0.98\textwidth]{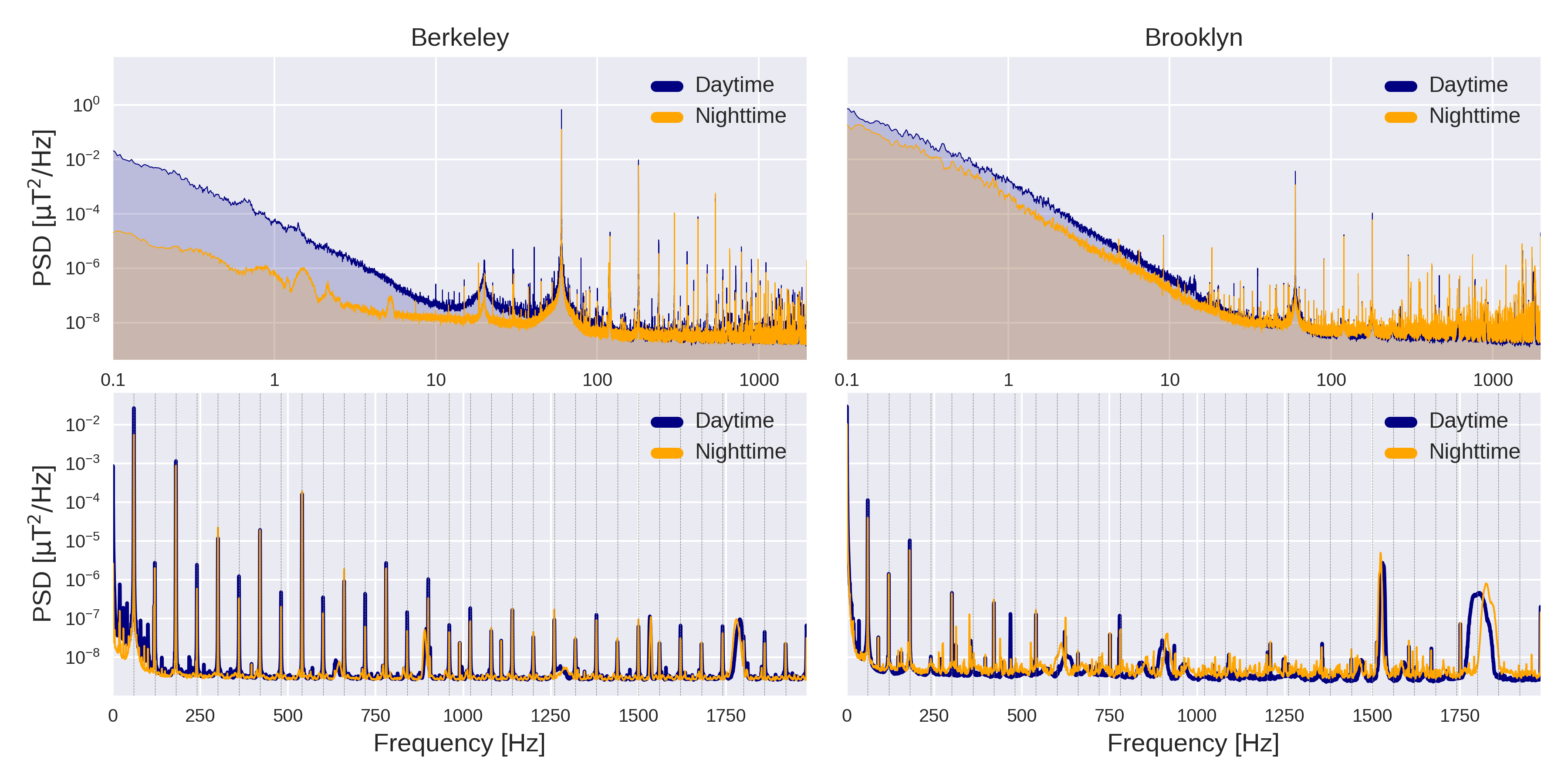}
\caption{High-frequency Power Spectral Density (PSD) from Berkeley (left) and Brooklyn (right) in both logarithmic (top) and linear (bottom) scales. The PSDs were produced using the full-rate data, sampled at 3,960\,Hz. The 60\,Hz power line and its harmonics can be seen clearly in the linear-scale PSD (bottom panel) and are highlighted by the thin dashed vertical lines.}
\label{Fig:psd_high_freq}
\end{figure*}

\begin{figure*}[t]
\includegraphics[width=0.98\textwidth]{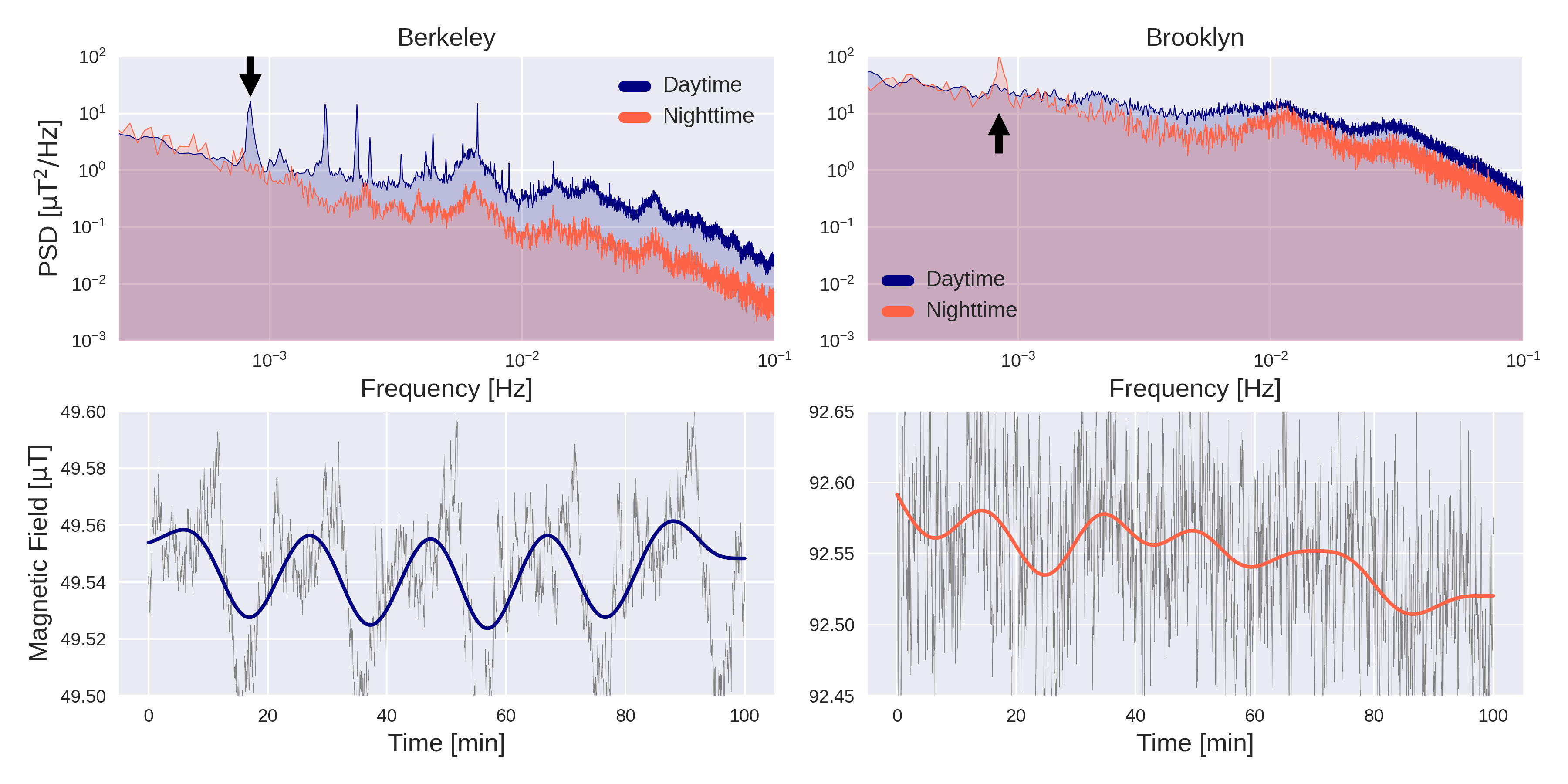}
\caption{Low-frequency PSDs with 20-minute signal extraction. The top panels show the logarithmic-scale PSDs produced using decimated data at 1\,Hz. The black arrows show a 20-minute periodic signal ($8.3\times10^{-4}\,\mathrm{Hz}$) that can be found in daytime variation from the Berkeley data and nighttime variations in Brooklyn. The bottom panels show the 20-minute periodic signal extracted from an ensemble average of 100-minute data regions after applying a high-pass filter with cutoff frequency at 0.001\,Hz.}
\label{Fig:psd_low_freq}
\end{figure*}

\begin{figure*}[t]
\includegraphics[width=0.98\textwidth]{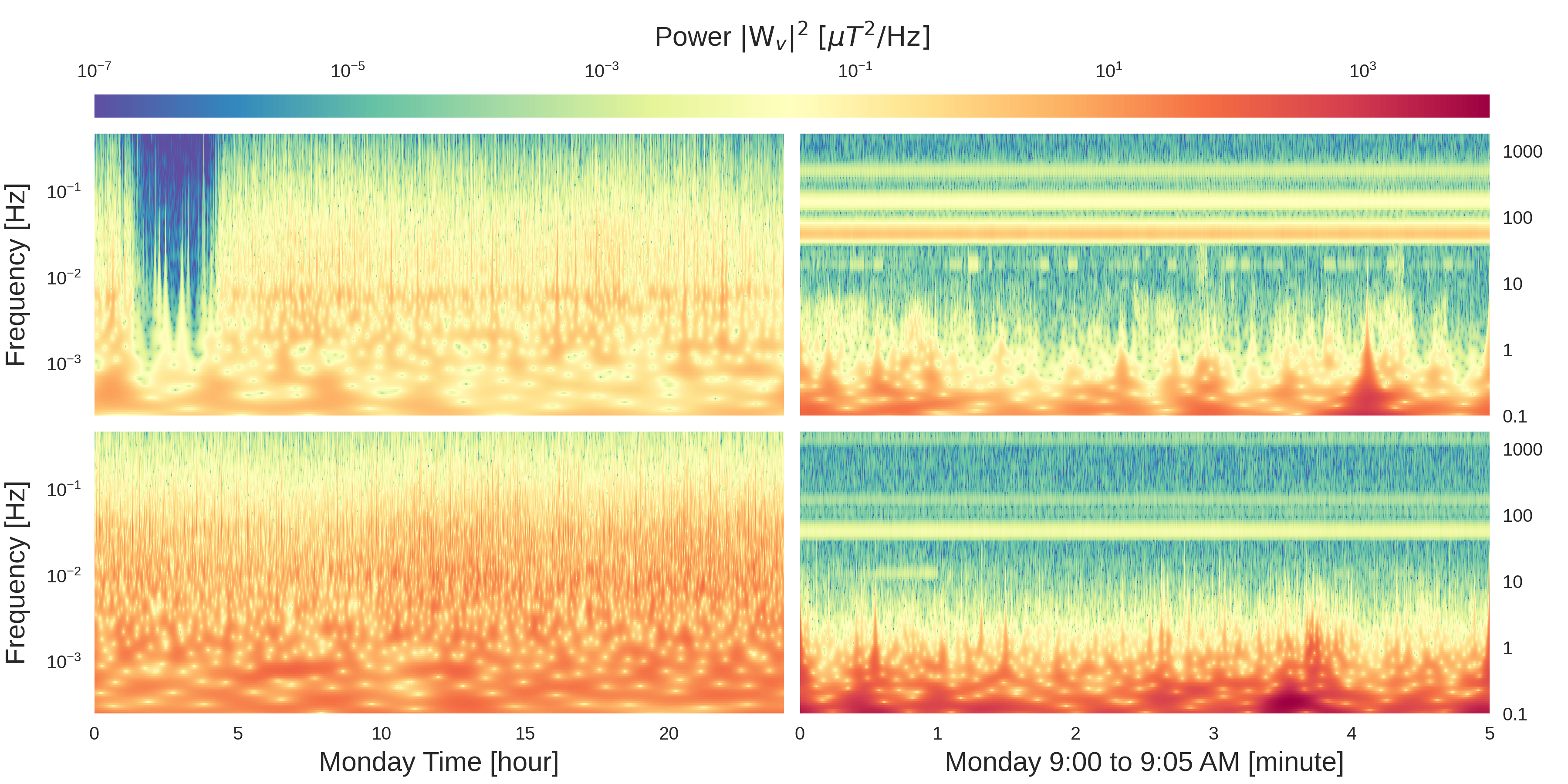}
\caption{Wavelet scalogram for a full day (left) and five minutes of data (right) for the first day (Monday) of both Berkeley (top) and Brooklyn (bottom) datasets. The full-day scalogram was achieved using the downsampled 1\,Hz data and plotted from the lowest available frequency, i.e. inverse of sampling rate to 500\,mHz. The 5-min scalograms were, on the other hand, produced using the full-rate data, thereby showing frequency content up to the Nyquist limit, i.e. half the sampling rate.}
\label{Fig:wavelet}
\end{figure*}

\begin{figure*}[t]
\includegraphics[width=0.98\textwidth]{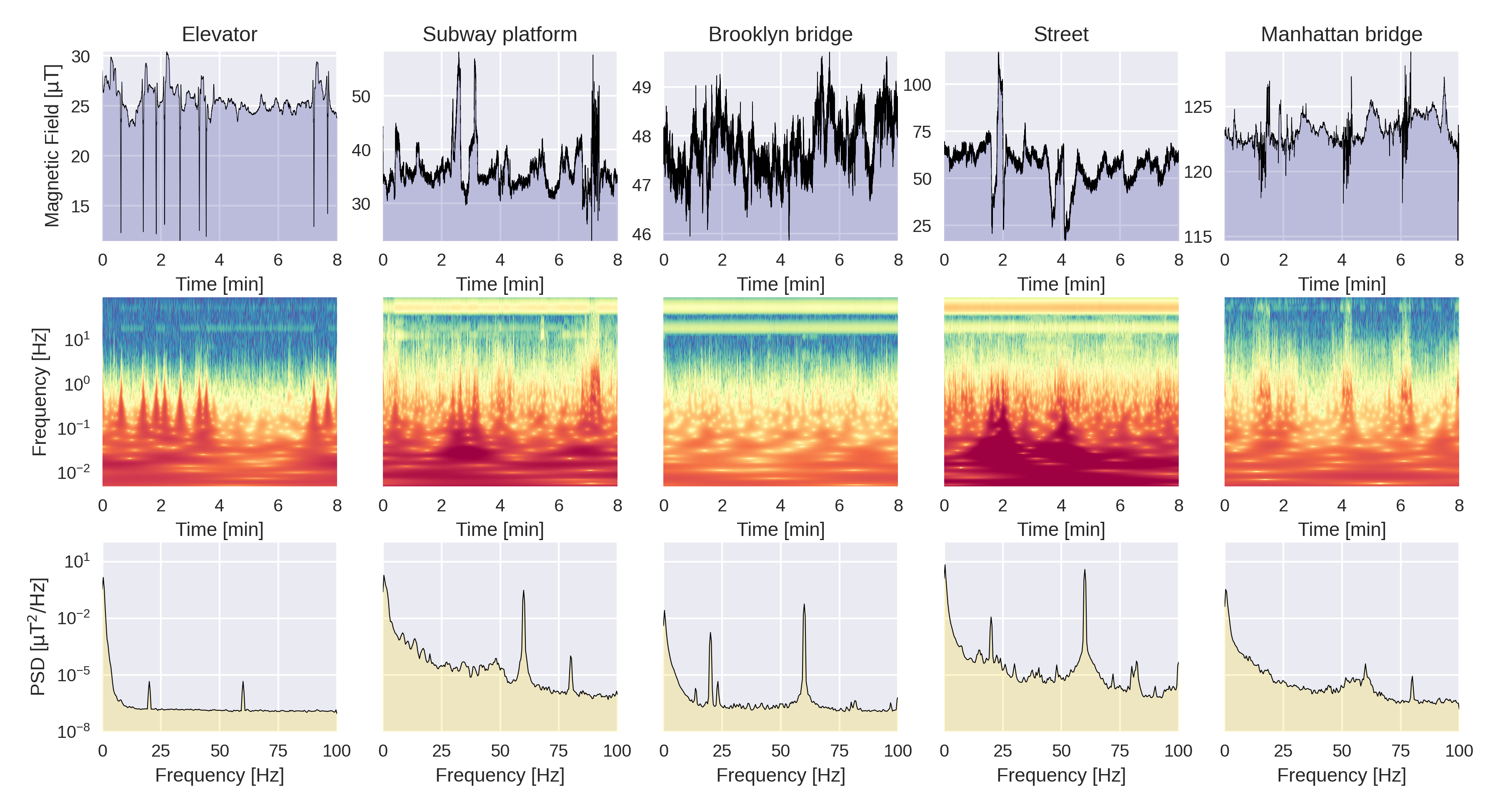}
\caption{Five samples of magnetic field time series at five different locations in Brooklyn. From left to right: (1) Elevator measurements were taken on the twelfth floor of Transit Building; (2) Subway measurements were acquired from the Jay Street Metro Tech station; (3) Brooklyn bridge measurements were taken underneath the bridge; (4) Street measurements were obtained on the sidewalk in front of the Transit Building in downtown Brooklyn; and (5) the Manhattan Bridge measurements were taken on top of the bridge from the middle of the walkway.}
\label{Fig:samples}
\end{figure*}

\begin{figure*}[t]
\includegraphics[width=0.98\textwidth]{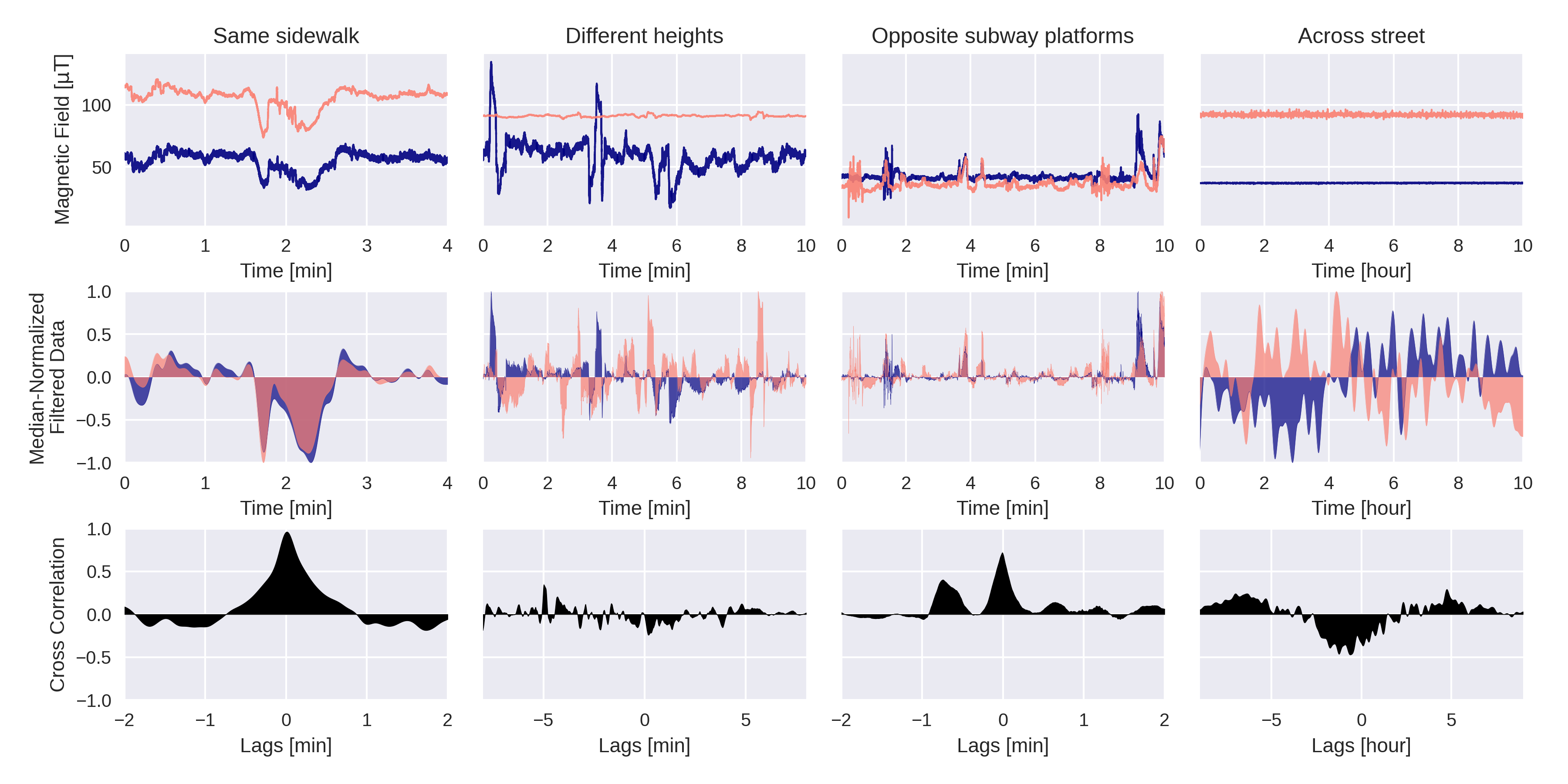}
\caption{Cross correlated data between two sensors. (1) Sensors placed seven meters apart from each other on the sidewalk; a low-pass filter with a cutoff frequency at 0.1\,Hz was applied to the data; (2) First sensor (blue) placed on the sidewalk and second sensor (red) in the CUSP office on the twelfth floor, low-pass filter applied with cutoff frequency at 10\,Hz; (3) Sensors placed at opposite ends of one of the platforms in the Jay Street - MetroTech subway station, low-pass filter applied with cutoff frequency at 10\,Hz; (4) Asynchronous measurements from 10 AM to 8 PM, one recorded data (blue) were streamed from the old CUSP office at One MetroTech Center on Monday 10 of October 2017 while the second set of measurements (red) were made from across the street in the Transit Building on Monday October 10, 2017, low-pass filter applied with cutoff frequency at 0.001\,Hz masking periodicity with timescale less than 16.7\,minutes.}
\label{Fig:xcorr}
\end{figure*}

\section*{Reproducibility Statement}

In an effort to make our work transparent and reproducible, the entirety of the data and codes used in this work have been made publicly available. Detailed instructions on how to access the data and reproduce the results presented here can be found in the online documentation available at \url{http://citymag.gitlab.io/nuri/paper}. The analysis was performed on 86\,GB worth of magnetic field data using the open-source  software \textsc{NURI} \cite{nuri_software}, specifically designed to analyze time-series data produced by our urban magnetometry network.

\section{Introduction}

Cities are among the most complex systems that are of utmost importance for humanity. The multifaceted and dynamic properties of cities are determined by intricate combinations of natural, anthropogenic, and socio-economic factors. In recent years, a novel approach to the study of the cities was introduced \cite{DOBLER2015115}, in which a city is studied, similar to an astronomical object in the multi-messenger astronomy approach, with an array of observational instruments such as, for example, multispectral cameras \cite{hyperspec1,hyperspec2,hyperspec3}. The analysis of such data has led to important insights into the working of cities \cite{s16122047,rs13081426,rs12162540}, of importance in such diverse areas as improving energy efficiency, reducing pollution, and increasing our understanding of social organization via the analysis of the work/sleep patterns of urban dwellers (with measurements carried out in a way to ensure privacy of individuals \cite{privacy}).

Motivated by the success of the multispectral approach, we built a prototype network for urban magnetometery \cite{2019_bowen} and conducted measurements in the San Francisco Bay Area, analyzing the dominant sources of magnetic signals and learning to extract subtle information in the presence of much larger backgrounds.

Here, we report the next step in the urban-magnetometry program, in which we compare the magnetic signatures of two cities, Berkeley (CA) and Brooklyn Borough of New York City (NY). Apart from the anticipated result that ``New York never sleeps'', our measurements indicate that each city has distinct magnetic signatures that can, perhaps, be exploited for the analysis of anomalies in city operation and long-term trends of the development of cities.

\section{Experimental details}

\subsection{Sensor type and data acquisition}

Two types of magnetometers were used to measure the magnetic field in Brooklyn. The base stations were built using three-axis fluxgate magnetometers manufactured by BioMed Jena GmbH. These Biomed sensors are tied to a specific location and have a sensitivity of about 70\,pT/$\sqrt{\mathrm{Hz}}$. A power supply from the same manufacturer is used to connect the magnetometers to a computer. Digitized magnetic field measurement data are transferred using a Universal Serial Bus (USB) connection. The data streamed from the Biomed sensors are sampled at 3960\,Hz and recorded on the computer using the publicly available {\sc UrbanMagnetometer} software\,\cite{data_acq}. Data from each magnetic field direction (that is, X, Y and Z) are stored hourly in separate binary files.

Field measurements were performed using Magnetoresistive Vector Magnetometers (VMR) manufactured by Twinleaf LLC\footnote{\url{https://twinleaf.com/vector/VMR/}} with sensitivity of 300\,pT/$\sqrt{\mathrm{Hz}}$. In this work, we analyze the total scalar field and not individual vector measurements from each axes.\footnote{There may be a small systematic effect in these measurements: the VMR sensors are based on HMC1001 series sensors and have variation over time and temperature such that the calibration of each axes can be off by as much as 10\,\%, which may cause the recorded ``scalar'' measurements to not be purely scalar.}  In terms of acquisition, the Twinleaf sensors do not require any data-acquisition device and can be powered directly from a laptop USB port, making them ideal for field measurements (see Section \ref{Sec:field}).

The geomagnetic field was acquired from the United States Geological Survey (USGS) using the open-source library Geomag Algorithms\footnote{\url{https://github.com/usgs/geomag-algorithms}}. The USGS station (FRN) nearest to Berkeley is located 200 miles away, in Fresno, California. For the Brooklyn data, the nearest USGS station (FRD) is in Corbin, Virginia, about 300 miles away from New York City.

\subsection{Activity period}

Data from the Biomed sensors were obtained over four weeks from each city during the calendar year 2016 for Berkeley and 2018 for Brooklyn. More specifically, the data used from Berkeley were taken from Monday March 14, 2016, through Monday April 11, 2016. The data from Brooklyn were acquired from Monday May 7, 2018, through Monday June 4, 2018; this period included the US Federal Memorial Day holiday, observed the last Monday of the month (05/28/2018) which is highlighted in red in Fig.\,\ref{Fig:full_time_series}. Holidays are usually characterized by a quieter magnetic environment due to a dropping off human activities; this can be particularly noticeable when the holiday falls on a weekday and the nearby environment surrounding the sensor has an overall magnetic field of higher amplitude during working hours. Finally, the geomagnetic field measured by the respective USGS station closest to each city has variations representing a small fraction of our measured variation, i.e. 3\% and below.

\subsection{Sensor locations}

The Berkeley measurements (originally presented in \citep{2019_bowen}) were conducted using geographically separated magnetometers in the city of Berkeley. The 4-week data used in this work were generated by one of the Biomed sensors located in a residential area 90 meters away from the Bay Area Rapid Transit (BART) rail system. The city of Berkeley has about 120,000 residents, living primarily in houses with a few low-rise buildings in the downtown. A BART line, which crosses the city, is the dominant source of magnetic field above the natural background during daytime \citep{fraser}.

In Brooklyn, the Biomed sensor was placed on the 12\textsuperscript{th} floor of the downtown-located Transit Building (370 Jay Street) in one of the corner offices of NYU's Center for Urban Science and Progress. In sharp contrast to Berkeley that has a population density of about 4,600 people per square kilometer (2020 census), Brooklyn is over 3 times denser with a population density approaching 15,000 people per square kilometer and its downtown constitutes a major transportation axis connecting it to downtown Manhattan. Located 40 meters underneath the sensor's position, beneath the Transit Building, is the Jay Street-MetroTech subway station, which is served by three subway lines at all times and by several additional lines during commute hours.

Due to limited resources, we were only able to stream data seamlessly from one base station in Brooklyn, located in the Jay Street building. As the observations are limited to a single-point measurement from the building, the question arises as to whether or not these magnetic field fluctuations are characteristic of the magnetic environment of Brooklyn, or, alternatively, if they are fluctuations in the magnetic field that are the result of a geographically localized set of magnetic sources in the Jay Street building. To address this concern, we performed a series of auxiliary experiments using multiple portable magnetic sensors (see Section \ref{Sec:field}) and demonstrate that while local processes measured by the Jay Street sensor are predominant, characteristic observations of the Brooklyn urban magnetic field can also be extracted from the data.

\section{Comparative Data Analysis}

\subsection{Time-domain observations}

The total scalar field for the entire four-week period for both cities is shown in Fig.\,\ref{Fig:full_time_series}. While daily variations of the magnetic field in Berkeley appear similar regardless of the day of the week (weekday fluctuations are similar to weekend fluctuations), the periodic behavior observed in the Brooklyn weekday data appears to stop on weekends and holidays. We note, however, that since the sensor in Brooklyn is placed within a business building, the drop in activities within the building during weekends and holidays (e.g., stopped elevators, lights off) represents a direct cause for the drop in magnetic field activities observed by the magnetometer. One should also note that the measured field is relatively far from the geomagnetic mean, indicating that the field on the sensor has a large contribution from a local source.

The change in variance during weekdays and weekends is shown on the top plots of Fig.\,\ref{Fig:distcomp}. We note that the dispersion of the magnetic field in Berkeley is two orders of magnitude less at night than during the day, dropping from $10^{-2}$ to $10^{-4}\,{\mathrm{\upmu T}^2}$, while nightly variations in downtown Brooklyn remains high with a variance lying above 0.1\,$\mathrm{\upmu T}^2$ all the time. Decreased amplitude fluctuations in Berkeley occur roughly between 1 to 4:30\,AM, when BART is not in service. We also note that nighttime activities differ slightly from weekday to weekend; that is a direct consequence of reduced public transport activities on weekends. In Brooklyn, however, the changes between daytime and nighttime variations are less pronounced and the weekend variation has only minor day/night variability. While a decrease in anthropogenic activity during weekdays is usually observed at around 4-7\,PM, when business activities are reduced, the decrease in the magnetic field only starts to be seen at around 11\,PM, thereby suggesting that the magnetic field measured by the Biomed sensor is not solely driven by the occupancy of the building.

The bottom plots of Fig.\,\ref{Fig:distcomp} show the distributions of magnetic field data for the full dataset as well as for daytime and nighttime periods. For each city, day and nighttime distributions were fitted independently using a skewed Gaussian profile
\begin{multline}
    f\left(x;A,\mu,\sigma,\gamma\right) = \frac{A}{\sigma\sqrt{2\pi}}\exp{\left[-(x-\mu)^2/\left(2\sigma^2\right)\right]}\\
    \times\left\{1+\mathrm{erf}\left[\frac{\gamma(x-\mu)}{\sigma\sqrt{2}}\right]\right\} ,
\end{multline}
where $A$, $\mu$, $\sigma$ and $\gamma$ correspond respectively to the amplitude, mean, standard deviation, and skewness of the profile, and $\mathrm{erf[\,]}$ is the error function. The best-fit results for each distribution are presented in Table\,\ref{table:best_fit}. Two observations can be made that distinguish the Berkeley magnetic field variations from Brooklyn. First, we note that while day and night time distributions recorded by the sensor in Brooklyn are centered around a consistent mean magnetic field of about 92.8\,$\mathrm{\upmu T}$, the mean of both distributions in Berkeley are different with high significance, from a mean of 49.361(6)\,$\mathrm{\upmu T}$ during the day to 49.925(9)\,$\mathrm{\upmu T}$ at night. The second observation that can be made is the change in skewness of the distribution in Berkeley where the nighttime distribution profile sees an increase in skewness compared with daytime variations. In Brooklyn, on the other hand, the distribution remains roughly Gaussian all the time with a low absolute skewness of around 1.

\begin{table}[t]
  \centering
  \renewcommand{\arraystretch}{1.7}
  \begin{tabular}{|l|c|c|c|c|}
    \hline
    \multirow{2}{*}{\textbf{Params}} &
    \multicolumn{2}{c|}{\textbf{Berkeley}} &
    \multicolumn{2}{c|}{\textbf{Brooklyn}}\\
    \cline{2-5} &
    \textbf{Daytime} &
    \textbf{Nighttime} & 
    \textbf{Daytime} & 
    \textbf{Nighttime}\\
    \hline
    $A$ & 67,980(407) & 11,488(654)  & 158,547(504) & 79,559(321) \\ \hline
    $\mu$ [$\mathrm{\upmu T}$]  & 49.361(6) & 49.925(9) & 92.802(14) & 92.833(18) \\ \hline
    $\sigma$ [$\mathrm{\upmu T}$] & 0.281(5) & 0.212(18) & 0.930(12) & 0.617(12) \\ \hline
    $\gamma$  & 1.39(8) & -4.61(148) & -1.15(5) & -0.93(7)\\ \hline
  \end{tabular}
  \caption{Best skewed Gaussian fit of daytime and nighttime magnetic field distributions for Berkeley and Brooklyn data.}
  \label{table:best_fit}
\end{table}

\subsection{Frequency content}

In Fig.\,\ref{Fig:psd_high_freq}, we show the Power Spectral Density (PSD) at high-frequency for both Berkeley and Brooklyn data. The drop in magnetic field activities in Berkeley at nighttime, identified in the variance plot (see top panels from Fig.\,\ref{Fig:distcomp}), can be explained by the decrease in low-frequency signals (up to 10\,Hz) in the PSD. We also note a significant difference in amplitude of the power-line and other high-frequency signals between both cities (see bottom panels from Fig.\,\ref{Fig:psd_high_freq}).

Low-frequency signals for both daytime and nighttime periods are shown in Fig.\,\ref{Fig:psd_low_freq}. We notice that a 20-minute signal at $8.3\times10^{-4}\,\mathrm{Hz}$ is observed during daytime in Berkeley which is known to be associated to the BART activities \cite{2019_bowen}. In Brooklyn, a similar signal is also observed, but at nighttime. In order to identify this 20-minute periodic signal in the time-series data, we made 100-minute averages of the daytime Berkeley and nighttime Brooklyn data. Applying a high-pass filter with a cutoff frequency at 0.001\,Hz allows us to improve the extraction and visibility of the 20-minute periodic signal (see bottom panels in Fig.\,\ref{Fig:psd_low_freq}). While the signal is already visible in the average 100-minute data for Berkeley, the noise in the unfiltered data from Brooklyn provides greater challenges to identifying the 20-minute signal.

In Fig.\,\ref{Fig:wavelet}, we show a scalogram which demonstrate the richness of urban magnetic field data. The quiet nighttime perturbations in Berkeley, previously shown in \cite{2019_bowen} are recovered. Using the full-rate data, one can see how high-frequency ranges are richer in anthropogenic activities. In particular, irregular signals below the power frequency can often be seen and are more prominently in the Berkeley data.

\subsection{Auxiliary Field Measurements}
\label{Sec:field}

The measurements previously made in Berkeley \cite{2019_bowen} revealed coherent magnetic field fluctuations in a geographically distributed magnetometer array. The
significant correlations between stations allowed identification of these
fluctuations with a ``global'' magnetic field which characterizes the
magnetic signature of the city of Berkeley (or, rather, the broader East
Bay). Therefore, in order to fully determine the signature of Brooklyn, or of New York City (NYC) at large, a comparative analysis of in-situ data taken from two different environments must be made.

In Fig.\,\ref{Fig:samples}, we show the behavior of the magnetic field in five distinct locations throughout Brooklyn. Each measurement was acquired using magnetoresistive vector magnetometers manufactured by Twinleaf LLC with data sampled at 200\,Hz and sensitivity of up to 300\,pT/$\sqrt{\mathrm{Hz}}$. As one can notice, downtown Brooklyn is an urban environment with a high diversity of magnetic field sources (e.g., elevators moving in buildings, cars on surface streets, subways crossing the Manhattan Bridge), thereby making the identification of a more global magnetic signature more challenging.

All field measurements were acquired using two stations to allow cross-correlation between both instruments. Figure\,\ref{Fig:xcorr} demonstrates how challenging the cross-correlation between individual stations that are geographically separated can be. While two stations placed close to each other (i.e., within a few meters, see first column) hold highly cross-correlated data, the information quickly becomes uncorrelated the further away one station is from another.

However, using low-pass filters, it becomes possible to correlate signals from different environments. For instance, in the second column of Fig.\,\ref{Fig:xcorr}, we cross-correlate the magnetic field recorded from the sidewalk in front of the Jay Street building with the magnetic field recorded from inside the twelfth floor of the Jay Street building and identified a correlated signal with a lag time of 5 minutes between both stations. Similarly, when recording the magnetic field from the inside of two buildings located across the street from each other (see last column in Fig.\,\ref{Fig:xcorr}), a noticeable anti-correlated behavior can be observed.

\section{Discussion}

\subsection{Optimal sensor network distributions}

This work represents an initial demonstration of the potential complexity in small-scale magnetic field variability in dense urban environments. Indeed, in the context of the ``magnetic field of a city'', our observations show that the power in high spatial frequency modes is larger in a dense city like Brooklyn; this may be a feature of larger cities with a wider variety of magnetic sources. Small-spatial-scale effects must therefore be considered when designing optimal sensor network systems so they can map out the spatial variability of magnetic field on multiple scales.

\subsection{Impact of small-scale effects on global field}

In this work, we attempt to characterize the global magnetic field of cities, that is, the portion of the magnetic field that has spatial and temporal variation, but is observed to have spatiotemporal correlations over the extent of the city system. A global magnetic system can be defined as the set of extended and point sources that contribute non negligibly to its magnetic field. In the case of a city system, this often contains multiple subsystems such as individual buildings and trains.

A subsystem generally contains a geographically localized (within a volume) set of magnetic sources, which are both extended and point-like in nature. For instance, building-specific fluctuations represent subsystems within larger systems where the time dependence includes effects from both the structure itself and the behavioral signals from the population that is using the structure. In our work, the base station located within the Jay Street building is subject to fields due to sources with a spatial extent comparable to that of the building, as well as any point-sources within its subsystem. Presumably, the field measured with a sensor in the building can also have contributions from other subsystems like trains, for example, from the subway station underneath the Jay Street building.

Spatial and temporal variations in the magnetic field are often observed in the data and can be due to a variety of reasons, including the motion of magnetized objects or time variations in the current generating the magnetic field. While not all point sources, that is, sources confined to a certain small volume, have their magnetic field measurable beyond the vicinity of a few nearby sensors, these sources come with some characteristic radial dependence that perturb the larger spatial scale magnetic field, thereby making the extraction of the underlying global city-system field more challenging to perform.

\subsection{Inferring global properties from local measurements}

Local measurements in a dense urban environment may have periodicities similar to the daily/weekly trends observed in all urban systems and one could probably argue that the subsystems are coupled to these large-scale systems. However, periodicities in the global field (e.g. the extended urban environment) are harder to measure.

We further point out that point-source perturbations can be used to understand buildings in a global field but from an in-situ experimental standpoint. However, it is hard to constrain any dynamics using a point-source assumption and a small number of sensors. Our interpretation of "subsystem" urban magnetic fields (i.e., local fluctuations) is that they basically consists of multiple dipoles (or multipoles) moving in potentially complex ways. A single sensor (even with vector measurements) is unable to uniquely determine a dipole moment/orientation. A minimum of two sensors are therefore needed; the same is true for fields generated by line current. An added challenge in understanding the variation of a localized source using a few measurement sites comes from the time dependence of the signals. 
Further investigations will need to address, to what extent localized fields can, in practice, be isolated and identified using a magnetometer network.

The magnetic environment within a subsystem may be highly chaotic. A determination of the large-scale field properties from local measurements
requires an analysis of the statistical variability in the measurements. An exception to a purely statistical approach might be when there is a single dominant source in the local subsystem which can be subtracted from the local field. 

\section{Conclusion}

In this pilot study, magnetic signatures obtained in different urban environments (Berkeley and Brooklyn) were compared. We find that there are major differences in magnetic signatures in these two test cases, for example, the difference in the contrast of magnetic signatures between day and night time.

There are many ways to analyze the rich urban magnetic data. As we have shown in this work, some of them allow reducing the complex data stream to a few key parameters that can be used to monitor the dynamics of the city.

The results of this work point towards a number of possible future directions. For example, this is an extension to sensor networks as introduced in \cite{2019_bowen}; correlation with other (nonmagnetic) data, for instance, those from multispectral cameras \cite{hyperspec3,s16122047}.  

A specific advantage of magnetometry for urban studies is that it can provide information on the functioning of infrastructure within its boundaries, but at a distance (e.g., a moving elevator or operating machinery within a building that can be detected from the outside), so uses might include, post-disaster assessment (e.g., vulnerability of partially destroyed buildings), infrastructure monitoring (e.g., assessment of sensorless bridges with short bursts of observations), monitoring the stability of the power grid (with instabilities being precursors of outages), etc. 

Some interesting multidisciplinary questions one could address include: How does an anomalous event such as epidemic or pandemic affect the urban magnetic signature? Are there significant monthly and/or seasonal variations of magnetic signatures? What are the origins of these variations, are they the same for different cities?, etc. It is the authors' belief that answering these questions of ``comparative urban magnetometry'' will teach us a lot about cities and this knowledge will eventually translate into tangible economic and social benefits.

There are also technical improvements that can benefit future urban-magnetometry studies. For example, if a measurement is done near a local source, vibrations of the sensor can lead to spurious signals. These, however, can be identified by correlating the magnetic readout with accelerometer data. In fact, the Twinleaf magnetometers that we used are already equipped with such auxiliary sensors. 

\nocite{*}
\bibliography{references}

\begin{thebibliography}{15}%
\makeatletter
\providecommand \@ifxundefined [1]{%
 \@ifx{#1\undefined}
}%
\providecommand \@ifnum [1]{%
 \ifnum #1\expandafter \@firstoftwo
 \else \expandafter \@secondoftwo
 \fi
}%
\providecommand \@ifx [1]{%
 \ifx #1\expandafter \@firstoftwo
 \else \expandafter \@secondoftwo
 \fi
}%
\providecommand \natexlab [1]{#1}%
\providecommand \enquote  [1]{``#1''}%
\providecommand \bibnamefont  [1]{#1}%
\providecommand \bibfnamefont [1]{#1}%
\providecommand \citenamefont [1]{#1}%
\providecommand \href@noop [0]{\@secondoftwo}%
\providecommand \href [0]{\begingroup \@sanitize@url \@href}%
\providecommand \@href[1]{\@@startlink{#1}\@@href}%
\providecommand \@@href[1]{\endgroup#1\@@endlink}%
\providecommand \@sanitize@url [0]{\catcode `\\12\catcode `\$12\catcode
  `\&12\catcode `\#12\catcode `\^12\catcode `\_12\catcode `\%12\relax}%
\providecommand \@@startlink[1]{}%
\providecommand \@@endlink[0]{}%
\providecommand \url  [0]{\begingroup\@sanitize@url \@url }%
\providecommand \@url [1]{\endgroup\@href {#1}{\urlprefix }}%
\providecommand \urlprefix  [0]{URL }%
\providecommand \Eprint [0]{\href }%
\providecommand \doibase [0]{http://dx.doi.org/}%
\providecommand \selectlanguage [0]{\@gobble}%
\providecommand \bibinfo  [0]{\@secondoftwo}%
\providecommand \bibfield  [0]{\@secondoftwo}%
\providecommand \translation [1]{[#1]}%
\providecommand \BibitemOpen [0]{}%
\providecommand \bibitemStop [0]{}%
\providecommand \bibitemNoStop [0]{.\EOS\space}%
\providecommand \EOS [0]{\spacefactor3000\relax}%
\providecommand \BibitemShut  [1]{\csname bibitem#1\endcsname}%
\let\auto@bib@innerbib\@empty
\bibitem [{\citenamefont {Dumont}(2022)}]{nuri_software}%
  \BibitemOpen
  \bibfield  {author} {\bibinfo {author} {\bibfnamefont {V.}~\bibnamefont
  {Dumont}},\ }\href {\doibase 10.5281/zenodo.5893775} {\enquote {\bibinfo
  {title} {{NURI: A Time-frequency and sensor network analysis software for
  urban magnetometry data.}}}\ } (\bibinfo {year} {2022})\BibitemShut {NoStop}%
\bibitem [{\citenamefont {Dobler}\ \emph {et~al.}(2015)\citenamefont {Dobler},
  \citenamefont {Ghandehari}, \citenamefont {Koonin}, \citenamefont {Nazari},
  \citenamefont {Patrinos}, \citenamefont {Sharma}, \citenamefont {Tafvizi},
  \citenamefont {Vo},\ and\ \citenamefont {Wurtele}}]{DOBLER2015115}%
  \BibitemOpen
  \bibfield  {author} {\bibinfo {author} {\bibfnamefont {G.}~\bibnamefont
  {Dobler}}, \bibinfo {author} {\bibfnamefont {M.}~\bibnamefont {Ghandehari}},
  \bibinfo {author} {\bibfnamefont {S.~E.}\ \bibnamefont {Koonin}}, \bibinfo
  {author} {\bibfnamefont {R.}~\bibnamefont {Nazari}}, \bibinfo {author}
  {\bibfnamefont {A.}~\bibnamefont {Patrinos}}, \bibinfo {author}
  {\bibfnamefont {M.~S.}\ \bibnamefont {Sharma}}, \bibinfo {author}
  {\bibfnamefont {A.}~\bibnamefont {Tafvizi}}, \bibinfo {author} {\bibfnamefont
  {H.~T.}\ \bibnamefont {Vo}}, \ and\ \bibinfo {author} {\bibfnamefont {J.~S.}\
  \bibnamefont {Wurtele}},\ }\bibfield  {title} {\enquote {\bibinfo {title}
  {Dynamics of the urban lightscape},}\ }\href {\doibase
  https://doi.org/10.1016/j.is.2015.06.002} {\bibfield  {journal} {\bibinfo
  {journal} {Information Systems}\ }\textbf {\bibinfo {volume} {54}},\ \bibinfo
  {pages} {115--126} (\bibinfo {year} {2015})}\BibitemShut {NoStop}%
\bibitem [{\citenamefont {Lebourgeois}\ \emph {et~al.}(2008)\citenamefont
  {Lebourgeois}, \citenamefont {Bégué}, \citenamefont {Labbé}, \citenamefont
  {Mallavan}, \citenamefont {Prévot},\ and\ \citenamefont
  {Roux}}]{hyperspec1}%
  \BibitemOpen
  \bibfield  {author} {\bibinfo {author} {\bibfnamefont {V.}~\bibnamefont
  {Lebourgeois}}, \bibinfo {author} {\bibfnamefont {A.}~\bibnamefont
  {Bégué}}, \bibinfo {author} {\bibfnamefont {S.}~\bibnamefont {Labbé}},
  \bibinfo {author} {\bibfnamefont {B.}~\bibnamefont {Mallavan}}, \bibinfo
  {author} {\bibfnamefont {L.}~\bibnamefont {Prévot}}, \ and\ \bibinfo
  {author} {\bibfnamefont {B.}~\bibnamefont {Roux}},\ }\bibfield  {title}
  {\enquote {\bibinfo {title} {Can commercial digital cameras be used as
  multispectral sensors? a crop monitoring test},}\ }\href {\doibase
  10.3390/s8117300} {\bibfield  {journal} {\bibinfo  {journal} {Sensors}\
  }\textbf {\bibinfo {volume} {8}},\ \bibinfo {pages} {7300--7322} (\bibinfo
  {year} {2008})}\BibitemShut {NoStop}%
\bibitem [{\citenamefont {Park}\ and\ \citenamefont
  {Crozier}(2013)}]{hyperspec2}%
  \BibitemOpen
  \bibfield  {author} {\bibinfo {author} {\bibfnamefont {H.}~\bibnamefont
  {Park}}\ and\ \bibinfo {author} {\bibfnamefont {K.~B.}\ \bibnamefont
  {Crozier}},\ }\bibfield  {title} {\enquote {\bibinfo {title} {Multispectral
  imaging with vertical silicon nanowires},}\ }\href {\doibase
  10.1038/srep02460} {\bibfield  {journal} {\bibinfo  {journal} {Scientific
  Reports}\ }\textbf {\bibinfo {volume} {3}},\ \bibinfo {pages} {2460}
  (\bibinfo {year} {2013})}\BibitemShut {NoStop}%
\bibitem [{\citenamefont {Duempelmann}, \citenamefont {Gallinet},\ and\
  \citenamefont {Novotny}(2017)}]{hyperspec3}%
  \BibitemOpen
  \bibfield  {author} {\bibinfo {author} {\bibfnamefont {L.}~\bibnamefont
  {Duempelmann}}, \bibinfo {author} {\bibfnamefont {B.}~\bibnamefont
  {Gallinet}}, \ and\ \bibinfo {author} {\bibfnamefont {L.}~\bibnamefont
  {Novotny}},\ }\bibfield  {title} {\enquote {\bibinfo {title} {Multispectral
  imaging with tunable plasmonic filters},}\ }\href {\doibase
  10.1021/acsphotonics.6b01003} {\bibfield  {journal} {\bibinfo  {journal} {ACS
  Photonics}\ }\textbf {\bibinfo {volume} {4}},\ \bibinfo {pages} {236--241}
  (\bibinfo {year} {2017})}\BibitemShut {NoStop}%
\bibitem [{\citenamefont {Dobler}\ \emph {et~al.}(2016)\citenamefont {Dobler},
  \citenamefont {Ghandehari}, \citenamefont {Koonin},\ and\ \citenamefont
  {Sharma}}]{s16122047}%
  \BibitemOpen
  \bibfield  {author} {\bibinfo {author} {\bibfnamefont {G.}~\bibnamefont
  {Dobler}}, \bibinfo {author} {\bibfnamefont {M.}~\bibnamefont {Ghandehari}},
  \bibinfo {author} {\bibfnamefont {S.~E.}\ \bibnamefont {Koonin}}, \ and\
  \bibinfo {author} {\bibfnamefont {M.~S.}\ \bibnamefont {Sharma}},\ }\bibfield
   {title} {\enquote {\bibinfo {title} {A hyperspectral survey of {N}ew {Y}ork
  {C}ity lighting technology},}\ }\href {\doibase 10.3390/s16122047} {\bibfield
   {journal} {\bibinfo  {journal} {Sensors}\ }\textbf {\bibinfo {volume} {16}}
  (\bibinfo {year} {2016}),\ 10.3390/s16122047}\BibitemShut {NoStop}%
\bibitem [{\citenamefont {Dobler}\ \emph {et~al.}(2021)\citenamefont {Dobler},
  \citenamefont {Bianco}, \citenamefont {Sharma}, \citenamefont {Karpf},
  \citenamefont {Baur}, \citenamefont {Ghandehari}, \citenamefont {Wurtele},\
  and\ \citenamefont {Koonin}}]{rs13081426}%
  \BibitemOpen
  \bibfield  {author} {\bibinfo {author} {\bibfnamefont {G.}~\bibnamefont
  {Dobler}}, \bibinfo {author} {\bibfnamefont {F.~B.}\ \bibnamefont {Bianco}},
  \bibinfo {author} {\bibfnamefont {M.~S.}\ \bibnamefont {Sharma}}, \bibinfo
  {author} {\bibfnamefont {A.}~\bibnamefont {Karpf}}, \bibinfo {author}
  {\bibfnamefont {J.}~\bibnamefont {Baur}}, \bibinfo {author} {\bibfnamefont
  {M.}~\bibnamefont {Ghandehari}}, \bibinfo {author} {\bibfnamefont
  {J.}~\bibnamefont {Wurtele}}, \ and\ \bibinfo {author} {\bibfnamefont
  {S.~E.}\ \bibnamefont {Koonin}},\ }\bibfield  {title} {\enquote {\bibinfo
  {title} {The urban observatory: A multi-modal imaging platform for the study
  of dynamics in complex urban systems},}\ }\href {\doibase 10.3390/rs13081426}
  {\bibfield  {journal} {\bibinfo  {journal} {Remote Sensing}\ }\textbf
  {\bibinfo {volume} {13}} (\bibinfo {year} {2021}),\
  10.3390/rs13081426}\BibitemShut {NoStop}%
\bibitem [{\citenamefont {Qamar}\ and\ \citenamefont
  {Dobler}(2020)}]{rs12162540}%
  \BibitemOpen
  \bibfield  {author} {\bibinfo {author} {\bibfnamefont {F.}~\bibnamefont
  {Qamar}}\ and\ \bibinfo {author} {\bibfnamefont {G.}~\bibnamefont {Dobler}},\
  }\bibfield  {title} {\enquote {\bibinfo {title} {Pixel-wise classification of
  high-resolution ground-based urban hyperspectral images with convolutional
  neural networks},}\ }\href {\doibase 10.3390/rs12162540} {\bibfield
  {journal} {\bibinfo  {journal} {Remote Sensing}\ }\textbf {\bibinfo {volume}
  {12}} (\bibinfo {year} {2020}),\ 10.3390/rs12162540}\BibitemShut {NoStop}%
\bibitem [{\citenamefont {Kontokosta}(2021)}]{privacy}%
  \BibitemOpen
  \bibfield  {author} {\bibinfo {author} {\bibfnamefont {C.~E.}\ \bibnamefont
  {Kontokosta}},\ }\bibfield  {title} {\enquote {\bibinfo {title} {Urban
  informatics in the science and practice of planning},}\ }\href {\doibase
  10.1177/0739456X18793716} {\bibfield  {journal} {\bibinfo  {journal} {Journal
  of Planning Education and Research}\ }\textbf {\bibinfo {volume} {41}},\
  \bibinfo {pages} {382--395} (\bibinfo {year} {2021})},\ \Eprint
  {http://arxiv.org/abs/https://doi.org/10.1177/0739456X18793716}
  {https://doi.org/10.1177/0739456X18793716} \BibitemShut {NoStop}%
\bibitem [{\citenamefont {Bowen}\ \emph {et~al.}(2019)\citenamefont {Bowen},
  \citenamefont {Zhivun}, \citenamefont {Wickenbrock}, \citenamefont {Dumont},
  \citenamefont {Bale}, \citenamefont {Pankow}, \citenamefont {Dobler},
  \citenamefont {Wurtele},\ and\ \citenamefont {Budker}}]{2019_bowen}%
  \BibitemOpen
  \bibfield  {author} {\bibinfo {author} {\bibfnamefont {T.~A.}\ \bibnamefont
  {Bowen}}, \bibinfo {author} {\bibfnamefont {E.}~\bibnamefont {Zhivun}},
  \bibinfo {author} {\bibfnamefont {A.}~\bibnamefont {Wickenbrock}}, \bibinfo
  {author} {\bibfnamefont {V.}~\bibnamefont {Dumont}}, \bibinfo {author}
  {\bibfnamefont {S.~D.}\ \bibnamefont {Bale}}, \bibinfo {author}
  {\bibfnamefont {C.}~\bibnamefont {Pankow}}, \bibinfo {author} {\bibfnamefont
  {G.}~\bibnamefont {Dobler}}, \bibinfo {author} {\bibfnamefont {J.~S.}\
  \bibnamefont {Wurtele}}, \ and\ \bibinfo {author} {\bibfnamefont
  {D.}~\bibnamefont {Budker}},\ }\bibfield  {title} {\enquote {\bibinfo {title}
  {A network of magnetometers for multi-scale urban science and informatics},}\
  }\href {\doibase 10.5194/gi-8-129-2019} {\bibfield  {journal} {\bibinfo
  {journal} {Geoscientific Instrumentation, Methods and Data Systems}\ }\textbf
  {\bibinfo {volume} {8}},\ \bibinfo {pages} {129--138} (\bibinfo {year}
  {2019})}\BibitemShut {NoStop}%
\bibitem [{\citenamefont {Zhivun}\ and\ \citenamefont
  {Dumont}(2018)}]{data_acq}%
  \BibitemOpen
  \bibfield  {author} {\bibinfo {author} {\bibfnamefont {L.}~\bibnamefont
  {Zhivun}}\ and\ \bibinfo {author} {\bibfnamefont {V.}~\bibnamefont
  {Dumont}},\ }\href {\doibase 10.5281/zenodo.1310088} {\enquote {\bibinfo
  {title} {Urban magnetometry data acquisition software},}\ } (\bibinfo {year}
  {2018})\BibitemShut {NoStop}%
\bibitem [{Note1()}]{Note1}%
  \BibitemOpen
  \bibinfo {note} {\protect \url
  {https://twinleaf.com/vector/VMR/}}\BibitemShut {NoStop}%
\bibitem [{Note2()}]{Note2}%
  \BibitemOpen
  \bibinfo {note} {There may be a small systematic effect in these
  measurements: the VMR sensors are based on HMC1001 series sensors and have
  variation over time and temperature such that the calibration of each axes
  can be off by as much as 10\protect \tmspace +\thinmuskip {.1667em}\%, which
  may cause the recorded ``scalar'' measurements to not be purely
  scalar.}\BibitemShut {Stop}%
\bibitem [{Note3()}]{Note3}%
  \BibitemOpen
  \bibinfo {note} {\protect \url
  {https://github.com/usgs/geomag-algorithms}}\BibitemShut {NoStop}%
\bibitem [{\citenamefont {Fraser-Smith}\ and\ \citenamefont
  {Coates}(1978)}]{fraser}%
  \BibitemOpen
  \bibfield  {author} {\bibinfo {author} {\bibfnamefont {A.~C.}\ \bibnamefont
  {Fraser-Smith}}\ and\ \bibinfo {author} {\bibfnamefont {D.~B.}\ \bibnamefont
  {Coates}},\ }\bibfield  {title} {\enquote {\bibinfo {title} {Large-amplitude
  {ULF} electromagnetic fields from {BART}},}\ }\href {\doibase
  https://doi.org/10.1029/RS013i004p00661} {\bibfield  {journal} {\bibinfo
  {journal} {Radio Science}\ }\textbf {\bibinfo {volume} {13}},\ \bibinfo
  {pages} {661--668} (\bibinfo {year} {1978})},\ \Eprint
  {http://arxiv.org/abs/https://doi.org/10.1029/RS013i004p00661}
  {https://doi.org/10.1029/RS013i004p00661} \BibitemShut {NoStop}%
\end{thebibliography}%

\end{document}